\documentclass[twocolumn,
preprintnumbers,prl,
superscriptaddress
]{revtex4-1}

\usepackage[height=8.85in,width=6.45in]{geometry}

\usepackage[utf8]{inputenc}
\usepackage[T1]{fontenc}
\usepackage{amsmath}
\usepackage{amssymb}
\usepackage{mathtools}
\allowdisplaybreaks

\usepackage{slashed}
\usepackage{braket}
\usepackage[usenames,dvipsnames,svgnames,table]{xcolor}
\usepackage[colorlinks,citecolor=DarkGreen,linkcolor=FireBrick,urlcolor=FireBrick,linktocpage,breaklinks=true]{hyperref}
\usepackage{graphicx}
\usepackage{tikz}

\usetikzlibrary{decorations.markings}
\tikzset{line/.style={line width=0.25mm},
curve/.style={line,smooth,tension=1},
->-/.style={decoration={
  markings,
  mark=at position #1 with {\arrow[>=stealth]{>}}},postaction={decorate}},
-<-/.style={decoration={
  markings,
  mark=at position #1 with {\arrow[>=stealth]{<}}},postaction={decorate}},
}

\usepackage{times}
\usepackage[scaled]{couriers}

\usepackage{bm}

\usepackage{xcolor}
\usepackage{mdframed}

\renewenvironment{figure}[1][]{
  \begin{originalfigure}[#1]
    \begin{mdframed}[linecolor=black!0,backgroundcolor=black!0]
}{
    \end{mdframed}
  \end{originalfigure}
}

\def\bZ{\mathbb{Z}}

\def\vvv{v}

\usepackage{colortbl}
\definecolor{llightyellow}{rgb}{1.0, 0.95, 0.7}
\definecolor{llightblue}{rgb}{0.7, 0.9, 1.0}
\definecolor{llightpink}{rgb}{1.0, 0.85, 0.95}
\definecolor{llightgreen}{rgb}{0.7, 1.0, 0.4}
\colorlet{lightyellow}{llightyellow!50!white}
\colorlet{lightblue}{llightblue!50!white}
\colorlet{lightgreen}{llightgreen!50!white}
\colorlet{lightpink}{llightpink!50!white}

\usepackage{listings}
\begin{document}

\preprint{CALT-TH-2021-034
}
\title{Numerical evidence for a Haagerup conformal field theory}

\author{Tzu-Chen Huang}
\affiliation{Walter Burke Institute for Theoretical Physics,
California Institute of Technology, Pasadena, CA 91125, USA}
\author{Ying-Hsuan Lin}
\affiliation{Jefferson Physical Laboratory, Harvard University, Cambridge, MA 02138, USA}
\author{Kantaro Ohmori}
\affiliation{Department of Physics, Faculty of Science, 
University of Tokyo, Bunkyo, Tokyo 113-0033, Japan}
\author{Yuji Tachikawa}
\affiliation{Kavli Institute for the Physics and Mathematics of the Universe (WPI), 
 University of Tokyo,  Kashiwa, Chiba 277-8583, Japan}
\author{Masaki Tezuka}
\affiliation{Department of Physics, Kyoto University,  Kyoto 606-8502, Japan}

\begin{abstract}
We numerically study an anyon chain based on the Haagerup fusion category, and find evidence that it leads in the long-distance limit to a conformal field theory whose central charge is $\sim 2$.  Fusion categories generalize the concept of finite group symmetries to non-invertible symmetry operations, and the Haagerup fusion category is the simplest one which comes neither from finite groups nor affine Lie algebras.  As such, ours is the first example of conformal field theories which have truly exotic generalized symmetries.
\end{abstract}

\maketitle
\textit{Introduction and summary.---}
Symmetry is one of the fundamental principles of physics, and is usually described by groups.
We can, however, envision physical systems whose symmetry is governed by mathematical concepts more general than groups.  
Indeed, quantum groups, which are certain deformations of Lie groups and are closely related to affine Lie algebras, are long known to describe various integrable models in 1+1 dimensions, such as the XXZ spin chain and generalizations.
In this letter, we consider systems with a related but different generalization of the concept of symmetry groups in 1+1 dimensions.
To motivate our particular generalization, we reinterpret a symmetry described by a finite group $G$ as specified by topological walls labeled by the elements $g\in G$ and implementing the symmetry operations.
Two such walls can be fused according to the group law, and the associativity axiom of the group allows them to be rearranged; see Fig.~\ref{fig:group}.

We can relax the requirement that the fusion of two walls $a$ and $b$ is given by a single wall, and instead allow the fusion to be given by a linear combination of walls, which we express using the fusion rule, 
\begin{equation}
a\times b = \bigoplus_c N^{c}_{ab} ~c, \label{fusionrule}
\end{equation}
with non-negative integers $N^c_{ab}$.
The associativity of walls is then expressed using the data known as F-symbols; see Fig.~\ref{fig:fusion}.
A symmetry described by a group can then be considered as a special case where $\sum_c N^{c}_{ab}=1$.

\begin{figure}[ht]
\centering
\begin{tikzpicture}[scale=.375]
\draw [line, ->-=.6] (0,-1.5) -- (0,1.5);
\draw (0,0.) node [left] {$g$};
\draw [line, ->-=.6] (1,-1.5) -- (1,1.5);
\draw (1,0) node [right] {$\vphantom{g}h$};
\draw [line,->] (2.5,0) -- (3.5,0);
\draw [line, ->-=.6] (4.5,-1.5) -- (4.5,1.5);
\draw (4.5,0) node [right] {$gh$};
\end{tikzpicture}
\qquad
\begin{tikzpicture}[scale=.25, baseline=0.2pt]
\draw [line, -<-=.5] (2,2) -- (2,-1) node[right] {$ghk$};
\draw [line, ->-=.4, ->-=1] (2,2) -- (0,4) node[above] {$g$};
\draw [line, ->-=1] (1,3) -- (2,4) node[above] {$\vphantom{g}h$};
\draw [line, ->-=1] (2,2) -- (4,4) node[above] {$\vphantom{g}k$};
\draw (2,1.9) node[left,shift={(-.5mm,0)}] {$gh$};
\draw [line,<->] (5,2) -- (7,2);
\draw [line, -<-=.5] (10,2) -- (10,-1) node[right] {$ghk$};
\draw [line, ->-=1] (10,2) -- (8,4) node[above] {$g$};
\draw [line, ->-=1] (11,3) -- (10,4) node[above] {$\vphantom{g}h$};
\draw [line, ->-=.4, ->-=1] (10,2) -- (12,4) node[above] {$\vphantom{g}k$};
\draw (10,1.9) node[right,shift={(.5mm,0)}] {$\vphantom{g}hk$};
\end{tikzpicture}
\caption{Finite group symmetries as realized by walls.\label{fig:group}}
\bigskip
\begin{tikzpicture}[scale=.375]
\draw [line, ->-=.6] (0,-1.5) -- (0,1.5);
\draw (0,0.) node [left] {$a\vphantom{b}$};
\draw [line, ->-=.6] (1,-1.5) -- (1,1.5);
\draw (1,0) node [right] {$b$};
\draw [line, ->-=1] (2.5,0) -- (3.5,0);
\draw (4,-0.4) node [right] {$\bigoplus\limits_{c} N_{ab}^c$};
\draw [line, ->-=.6] (8.5,-1.5) -- (8.5,1.5);
\draw (8.5,0) node [right] {$\vphantom{b}c$};
\end{tikzpicture}
\begin{tikzpicture}[scale=.25]
\draw [line, -<-=.5] (2,2) -- (2,0) node[right] {$d$};
\draw [line, ->-=.4, ->-=1] (2,2) -- (0,4) node[above] {$a$};
\draw [line, ->-=1] (1,3) -- (2,4) node[above] {$b$};
\draw [line, ->-=1] (2,2) -- (4,4) node[above] {$c$};
\draw (2,1.9) node[left,shift={(-.5mm,0)}] {$e$};
\draw (5,2) node[right] {$=\sum\limits_{f} \left(F_{d}^{abc}\right)_{ef}$};
\draw [line, -<-=.5] (18,2) -- (18,0) node[right] {$d$};
\draw [line, ->-=1] (18,2) -- (16,4) node[above] {$a$};
\draw [line, ->-=1] (19,3) -- (18,4) node[above] {$b$};
\draw [line, ->-=.4, ->-=1] (18,2) -- (20,4) node[above] {$c$};
\draw (18,1.9) node[right] {$f$};
\end{tikzpicture}
\caption{Generalized symmetries described by fusion categories.\label{fig:fusion}}
%
\end{figure}

Anyons in 2+1 dimensional systems are described using a similar set of data.
The difference is that anyons require one additional piece of data, known as braiding, to express how two anyons can be exchanged. 
Generalized symmetries in 1+1 dimensions are mathematically formalized using fusion categories
\footnote{%
The use of fusion categories in this context goes back to e.g.\ \cite{Fuchs:2002cm,Carqueville:2012dk}, and was popularized more recently in e.g.\ \cite{Bhardwaj:2017xup,Chang:2018iay}. 
The standard mathematical reference is \cite{etingof2016tensor}.
}, whereas anyons in 2+1 dimensions are described using fusion categories equipped with non-degenerate braiding, also known as modular tensor categories
\footnote{%
Modular tensor categories were first described in \cite{Moore:1988qv}, although the mathematical structure was not named as such. 
An important, relatively recent reference on the condensed-matter side was \cite{Kitaev:2005hzj}.
}.
In particular, the fusion rule \eqref{fusionrule}
can be non-commutative in the former, whereas it is commutative in the latter.

All these would be empty words if there were no systems with such generalized symmetries. 
It is known that any rational conformal field theory has part of its generalized symmetries given by modular tensor categories \cite{Verlinde:1988sn,Moore:1988qv}, 
which are closely related to quantum groups and affine Lie algebras.
While almost all known fusion categories are related to either finite groups, quantum groups, 
or  affine Lie algebras
\footnote{More precisely, what we mean is that almost all concretely known fusion categories are categorically Morita-equivalent to either pointed fusion categories associated to finite groups or modular tensor categories associated to quantum groups and affine Lie algebras.
}, 
there are exceptions, the simplest among which is the Haagerup fusion category \cite{HaagerupOriginal,Grossman_2012}.
It turns out \cite{Buican:2017rxc,Aasen:2020jwb} that the famous anyon chain \cite{Feiguin:2006ydp}, originally defined with modular tensor categories as the input, makes perfect sense with fusion categories, and 
has the input fusion category as a generalized symmetry
\footnote{%
Calling the resulting chain as an `anyon' chain is a misnomer, since fusion categories which are not modular tensor categories do not describe anyons.
We still use the terminology `anyon chain' 
for the lack of a better name.
}.
The nature of the ground state, however, is a dynamical problem,
and it is of great interest to study whether there are conformal fixed points that do not spontaneously break the Haagerup symmetry.
The main objective of this letter is to announce that we obtained  numerical evidence that the anyon chain based on the Haagerup fusion category becomes a conformal field theory in the long distance limit.
The central charge was found to be $\sim 2$.
The details will be presented elsewhere.

Before proceeding, we note that almost the same model was considered previously \cite{Wolf:2021kkq}, and that its conformality will be announced by another group \cite{Verstraete} simultaneously with the present work.

\medskip
\textit{The model and the methods.---}
We start from the Haagerup fusion category $H_3$
\footnote{
What Haagerup originally constructed in \cite{HaagerupOriginal} was a subfactor, from which two fusion categories categorically Morita-equivalent to each other could be extracted. 
In \cite{Grossman_2012}, all fusion categories categorically Morita-equivalent to the two fusion categories coming from the Haagerup subfactor were identified.
The fusion category used here, $H_3$, was one of them.}
which has six labels $1$, $a$, $a^2$, $\rho$, $a\rho$, $a^2\rho$, and the fusion rule
\begin{equation}
\label{frh}
a^3=1, \quad
a\rho=\rho a^{-1},\quad
\rho^2 = 1+ (1+a+a^2)\rho.
\end{equation}
Note that this is non-commutative, and therefore $H_3$ does not admit any braiding.
The F-symbols were found in \cite{Titsworth,osborne2019fsymbols,Huang:2020lox}.
The anyon chain \cite{Feiguin:2006ydp} has basis states given by the diagram
\begin{equation}
    \label{Basis}
\ket{a_1 a_2 \cdots a_L} =
\begin{gathered}
\begin{tikzpicture}[scale=.8]
\draw [line,-<-=.1,-<-=.3,-<-=.7,-<-=.9] (-1,0) node {$/\!/~$} -- (-.5,0) node [below] {$a_1$} -- (.5,0) node [below] {$a_2$}
-- (1.5,0) node [below] {$\dotsb$}
-- (2.5,0) node [below] {$a_L$} -- (3.5,0) node [below] {$a_1$} -- (4,0) node {$~/\!/$};
\draw [line,-<-=.5] (0,1) node [above] {$\rho$} -- (0,0);
\draw [line,-<-=.5] (1,1) node [above] {$\rho$} -- (1,0);
\draw [line,-<-=.5] (2,1) node [above] {$\rho$} -- (2,0);
\draw [line,-<-=.5] (3,1) node [above] {$\rho$} -- (3,0);
\end{tikzpicture}
\end{gathered}
\end{equation}
where the fusion $a_i \times \rho$ contains $a_{i+1}$, and the slashes $/\!/$ represent periodic identification.  
For each adjacent pair of $\rho$ attached to $a_{i-1}$, $a_i$, $a_{i+1}$,
we can define an operator $P_c^{(i)}$ projecting the fusion of the pair of $\rho$'s to the label $c$.  Explicitly,
\begin{multline}
P^{(i)}_c \ket{a_{i-1} a_{i} a_{i+1}} =\\
\sum_{a_i'}
(F_{a_{i+1}}^{a_{i-1} \rho\rho})_{a_i c}
\overline{(F_{a_{i+1}}^{a_{i-1} \rho\rho})}_{{a_i'} c}
\ket{a_{i-1} a'_{i} a_{i+1}} \, ,
\end{multline}
where we assume that the F-symbols have been chosen to be unitary.  The Hamiltonian considered is given by the `ferromagnetic' pure $\rho$ projector, i.e.
\begin{equation}
H = - \sum_i P_\rho^{(i)}.
\end{equation}

We studied this spin chain with two numerical techniques: finite-system density matrix renormalization group (DMRG) \cite{White:1992zz} using iTensor \cite{Fishman:2020gel} up to $L=36$ (periodic) and $L=144$ (open), and exact diagonalization up to $L=18$ (periodic)
\footnote{DMRG was performed under maximum bond dimension 1600, truncation error cutoff $10^{-8}$, and energy$\times L$ accuracy $10^{-2}$.  We imposed the fusion rule by introducing penalty terms with coefficient $U = 2$, but also found the low-lying spectra to be insensitive to $U$.
}.

Our Hamiltonian is parity-symmetric,
and has an obvious $\bZ_3$ symmetry rotating the labels as $1\to a\to a^2\to 1$ and $\rho\to a\rho\to a^2\rho\to \rho$.
Additionally, our numerical results indicate that the nature of the ground state of the Hamiltonian depends strongly on  $L$ mod 3, signaling that the shift by one lattice site generates an additional $\bZ_3$ internal symmetry in the continuum limit, 
akin to the case of the $SU(n)$ Heisenberg antiferromagnetic chain with $n$ states per site,
where the lattice shift generates a $\bZ_n$ symmetry in the continuum limit \cite{Affleck:1988wz}.
Hence, $L$ not divisible by 3 is to be interpreted as shift-$\bZ_3$-twisted Hilbert spaces.
We also note that this shift-$\bZ_3$ symmetry commutes with the Haagerup symmetry, since it comes from a lattice translation.
When we studied open chains, we used Dirichlet boundary conditions $a_1 = \rho$, breaking both $\bZ_3$ symmetries.

\begin{figure*}
    \centering
    \footnotesize
    \begin{tabular}{cc|cc}
    \multicolumn{2}{c|}{Periodic} & \multicolumn{2}{c}{Open}  \\
    \hline
    \tiny
    $\begin{array}{l@{\,}ll@{\,}l}
        \alpha &= -0.84521(2) \\
        \beta &= -0.0049(8)&
        \gamma &= -1.10(1)  \\
    \end{array}$
    & \tiny
    $\begin{array}{l@{\,}ll@{\,}l}
        \alpha &= 0.0001(3) \\
        \beta &= 0.00(2)   &
        \gamma &= 1.5(2)\\
        \end{array}$
    & \tiny
    $\begin{array}{l@{\,}ll@{\,}l}
        \alpha &= -0.845339(3) \\
        \beta &= 0.59406(5) &
        \gamma &= -0.249(1) 
        \end{array}$
    & \tiny
    $\begin{array}{l@{\,}ll@{\,}l}
        \alpha &= -0.000028(6) \\
        \beta &= 0.0052(9) &
        \gamma& = 1.84(2) 
        \end{array}$
    \\
    \includegraphics[width=0.24\textwidth]{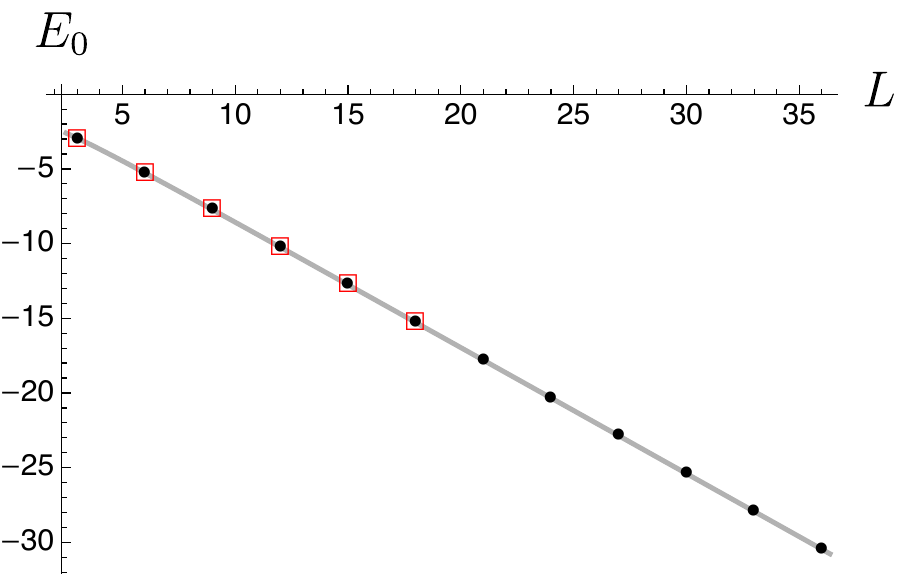} &
    \includegraphics[width=0.24\textwidth]{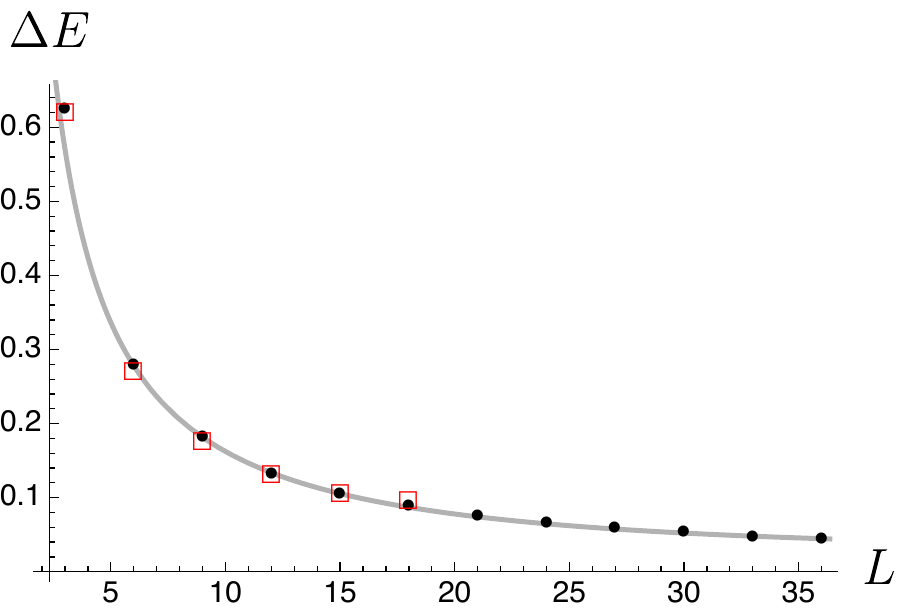}& 
    \includegraphics[width=0.24\textwidth]{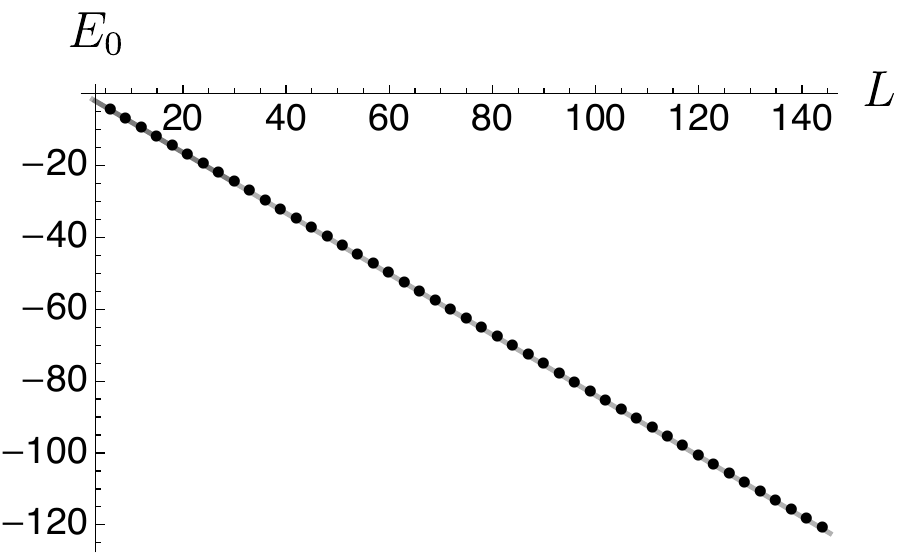}&
    \includegraphics[width=0.24\textwidth]{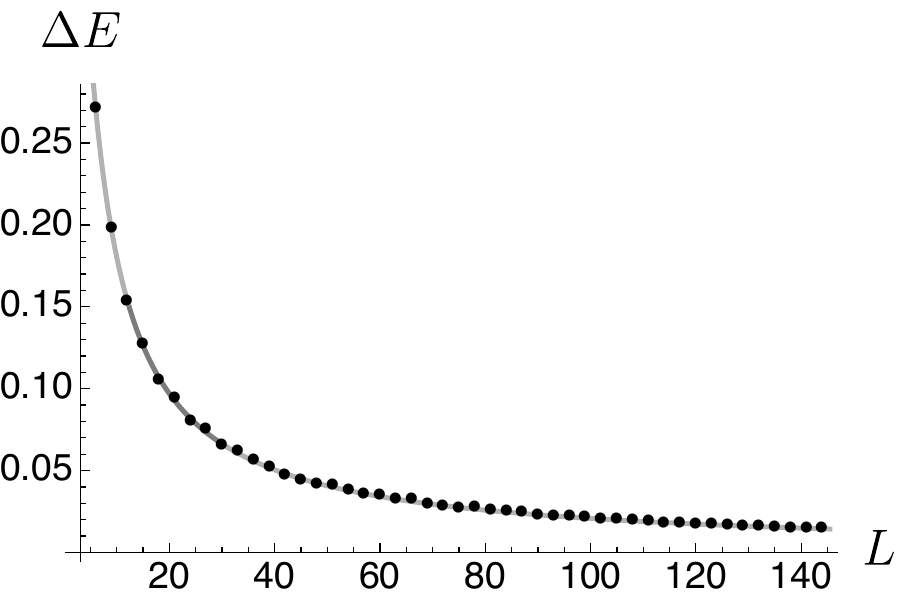}
    \end{tabular}
        \caption{Ground state energy $E_0$ and gap $\Delta E$ as functions of chain length $L$, juxtaposed with fits to \eqref{GSFit}. Black dots: DMRG; Red squares: exact diagonalization. The errors in the parameters shown here only reflect the fitting errors.
        }
        \label{fig:energy}
\end{figure*}

\medskip
\textit{The results.---}
A state $|\psi_i\rangle$ in the continuum conformal field theory with scaling dimension $\Delta_i$ is expected to manifest on quantum spin chains of varying length $L$ as states with energies
\begin{equation}
    E_i = 
    \begin{cases}
    \displaystyle
    \alpha L +  \frac{\vvv}L(\Delta_i - \frac{c}{12}) + o\left(\frac{1}{L} \right)  , \\
    \displaystyle
    \alpha L + \beta' + \frac{\vvv}{2L}(\Delta_i - \frac{c}{24}) + o\left(\frac{1}{L} \right)  , \\
    \end{cases}
    \label{generalFit}
\end{equation}
where the first/second line is for the periodic/open chain.
The coefficient $\alpha$ is the density of energy per site,
the constant term $\beta'$ captures the energy contribution from boundaries,
and the coefficient $\vvv$ fixes the `speed of light' in lattice units.
Let $E_0$ be the ground state energy, $E_1$ be the energy of the first excited state neutral under the intrinsic $\mathbb{Z}_3$, and $\Delta E = E_1-E_0$ be the gap.  
The data and fit of $E_0$ and $\Delta E$ to the ansatz 
\begin{equation}
    \label{GSFit}
        \alpha L + \beta + \frac{\gamma}{L} 
    \end{equation}
are shown in Fig.~\ref{fig:energy}
\footnote{
We intentionally did not include the subleading corrections to \protect\eqref{GSFit} when performing the fit,
since their form would depend strongly on the spectra and the operator product expansion coefficients.
To ameliorate the situation, we only used the values in $L \ge 18$ for the fit, hoping that the subleading corrections would be smaller there.
We also note that the fitting errors shown in Fig.~\ref{fig:energy} constitute small fractions of the overall errors in the parameters.  The DMRG energy$\times L$ accuracy setting of $10^{-2}$ limits $\gamma$ to $10^{-2}$ accuracy, and the maximum bond dimension and truncation error cutoff introduce errors that are more difficult to quantify.  Note that the ratio between periodic and open of the fitted values of $\gamma$ for $E_0$ is $1.10/0.249 \sim 4.4$, which when compared to the theoretical value 4 means that there is at least a combined 10\% error present.
}.
Note that DMRG and exact diagonalization agree for overlapping values of $L$.

\begin{figure}[!b]
\footnotesize
    \begin{tabular}{l|l}
    \multicolumn{1}{c|}{Periodic} & \multicolumn{1}{c}{Open}
    \\
    \hline
     $\log\Delta E =$ 
    &
     $\log\Delta E = $
    \\
     \qquad $-1.01(2) \log L + 0.49(8)$ &
     \qquad $-1.0(2) \log L + 0.73(9)$
    \\
    \includegraphics[width=0.5\textwidth]{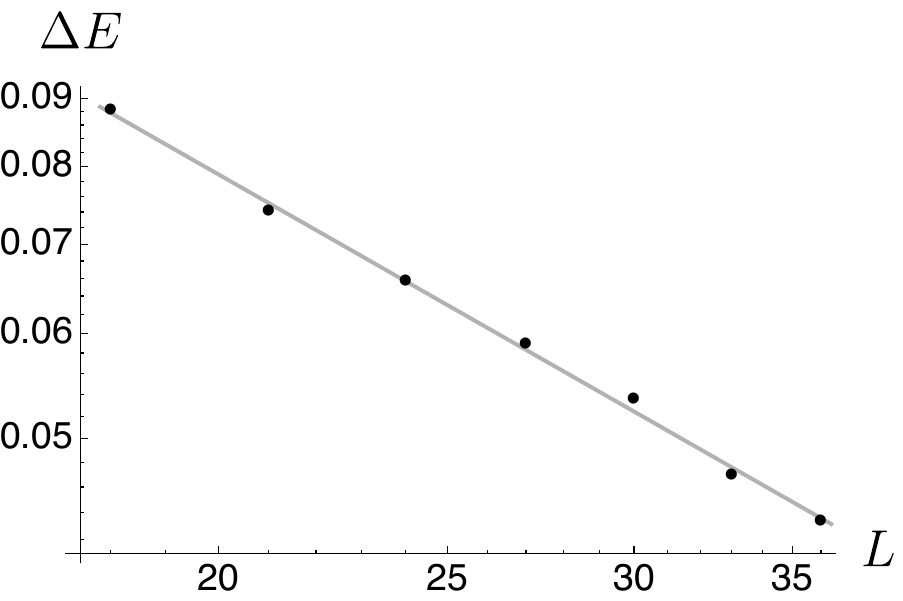}
    &
    \includegraphics[width=0.5\textwidth]{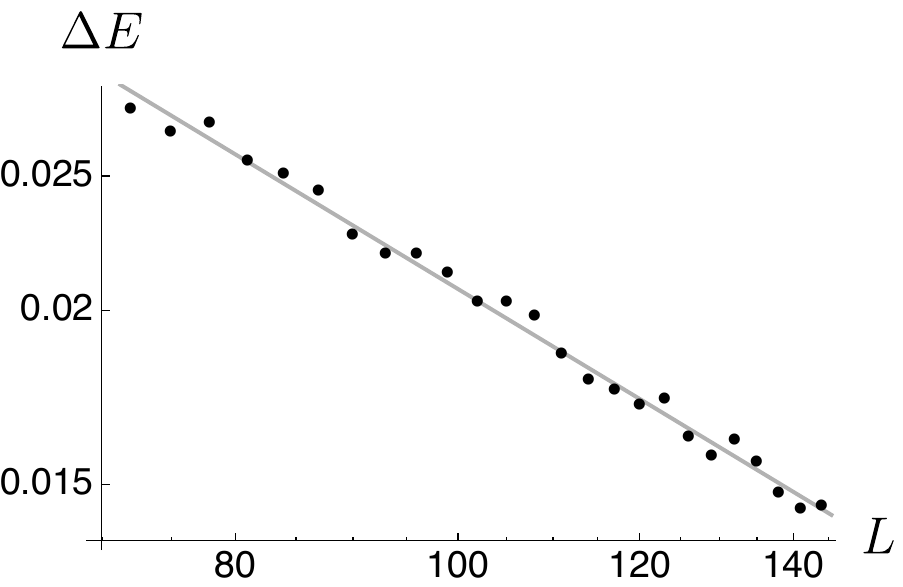}
    \end{tabular}
    \caption{Scaling of the gap with the system size.  Fits of $\log \Delta E$ over $\log L$ show that the dynamical exponent is $z \sim 1$.}
    \label{fig:puredyn}

\end{figure}

The $L$ dependence of $\Delta E$ can also be described 
in terms of the dynamical exponent, $\Delta E=L^{-z}$.
For a conformal field theory, $z$ should be $1$.
In Fig.~\ref{fig:puredyn} we exhibit the scaling of the gap with the system size.
Our linear fits show that $z \sim 1$.

\begin{figure}[!b]
    \begin{tabular}{c|c}
    \footnotesize Periodic & \footnotesize Open
    \\
    \hline
    \footnotesize $c=2.034(4)$ & \footnotesize $c=2.11(7)$
    \\
    \includegraphics[width=0.5\textwidth]{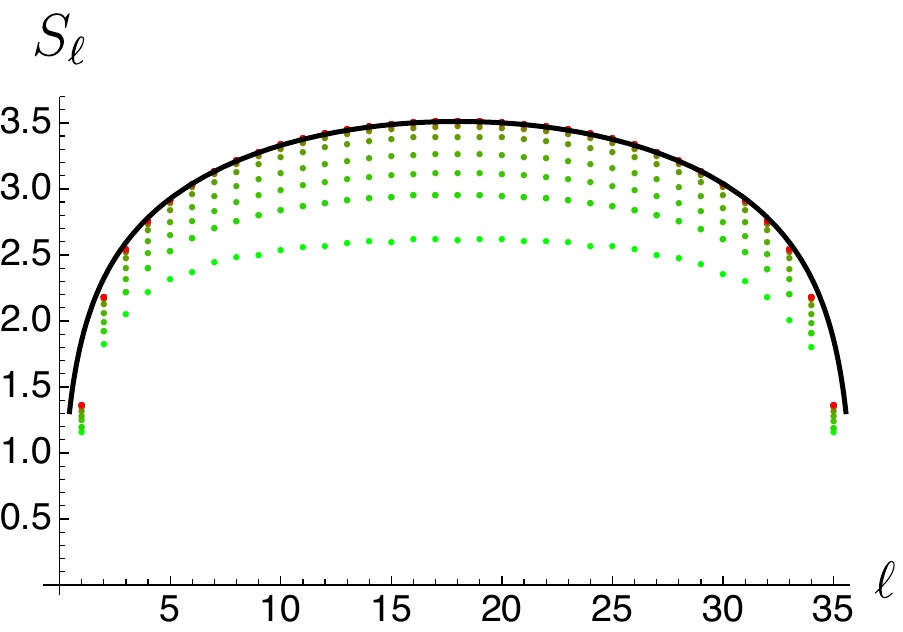}
    &
    \includegraphics[width=0.5\textwidth]{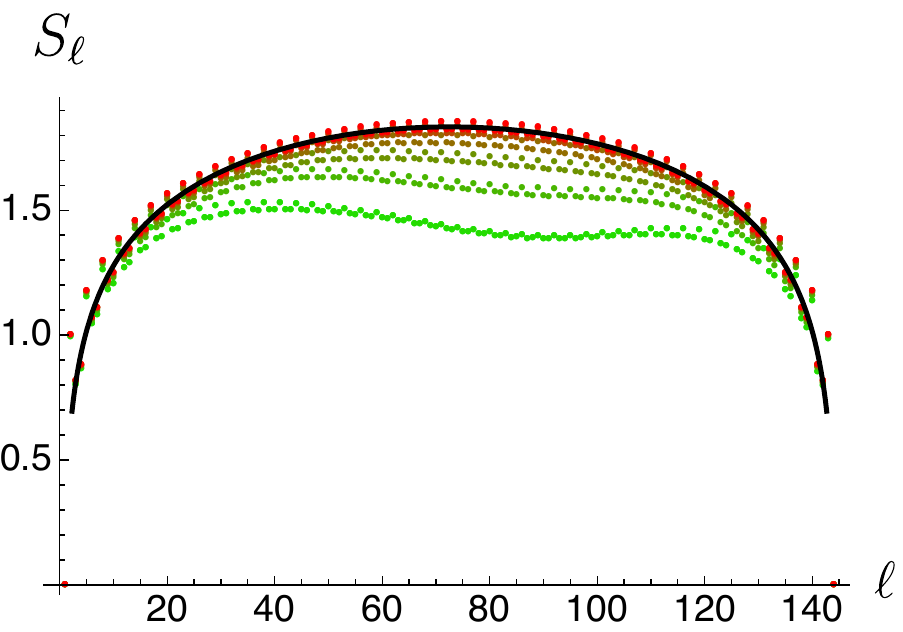}
    \end{tabular}
    \caption{Entanglement entropy curve of the ground state and fit to \eqref{entanglement}.  
    The color scheme shows the evolution over DMRG sweeps.
    }
    \label{fig:ee}

\end{figure}

Next, we computed the entanglement entropy in the ground state for varying interval length $\ell$, and fit to
\begin{equation}\begin{aligned}
\label{entanglement}
    S_\ell &= \begin{cases}
    \displaystyle
    \frac{c}{3} \log\left( \frac{L}{\pi} \sin\frac{\pi \ell}{L} \right) + \text{const} \, , 
    \\
    \displaystyle
    \frac{c}{6} \log\left( \frac{L}{\pi} \sin\frac{\pi (\ell-\frac12)}{L} \right) + \text{const} \, , 
    \end{cases}
\end{aligned}\end{equation}
where the upper/lower equation is for the periodic/open chain.
Upon discarding $\ell < \frac{L}{5}$ and $\ell > \frac{4L}{5}$ from the fit, the results for $L=36$ (periodic) and $L=144$ (open) are shown in Fig.~\ref{fig:ee}, where the central charge is estimated to be $c=2.0(1)$.

\newpage

\begin{figure*}[ht]
\centering
    \includegraphics[height=0.175\textheight]{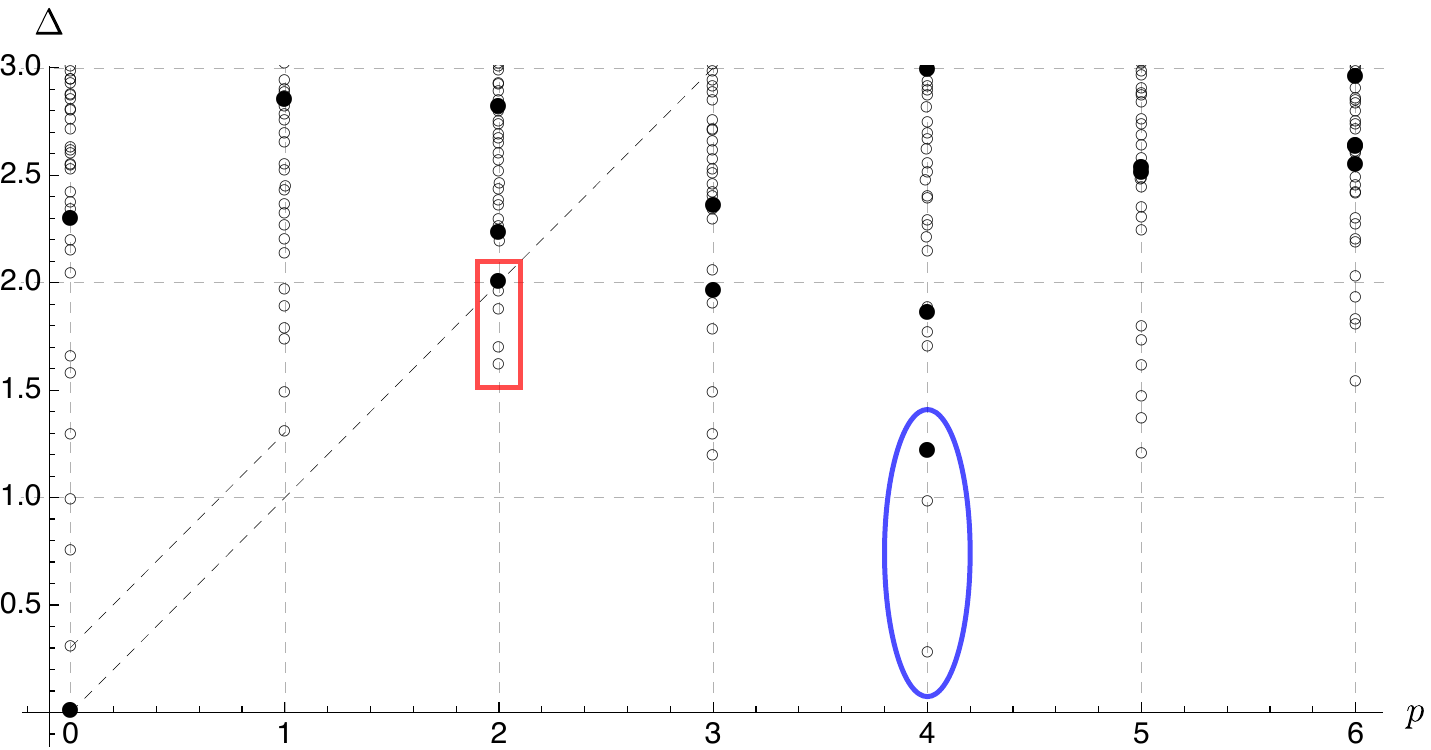}
    \includegraphics[height=0.175\textheight]{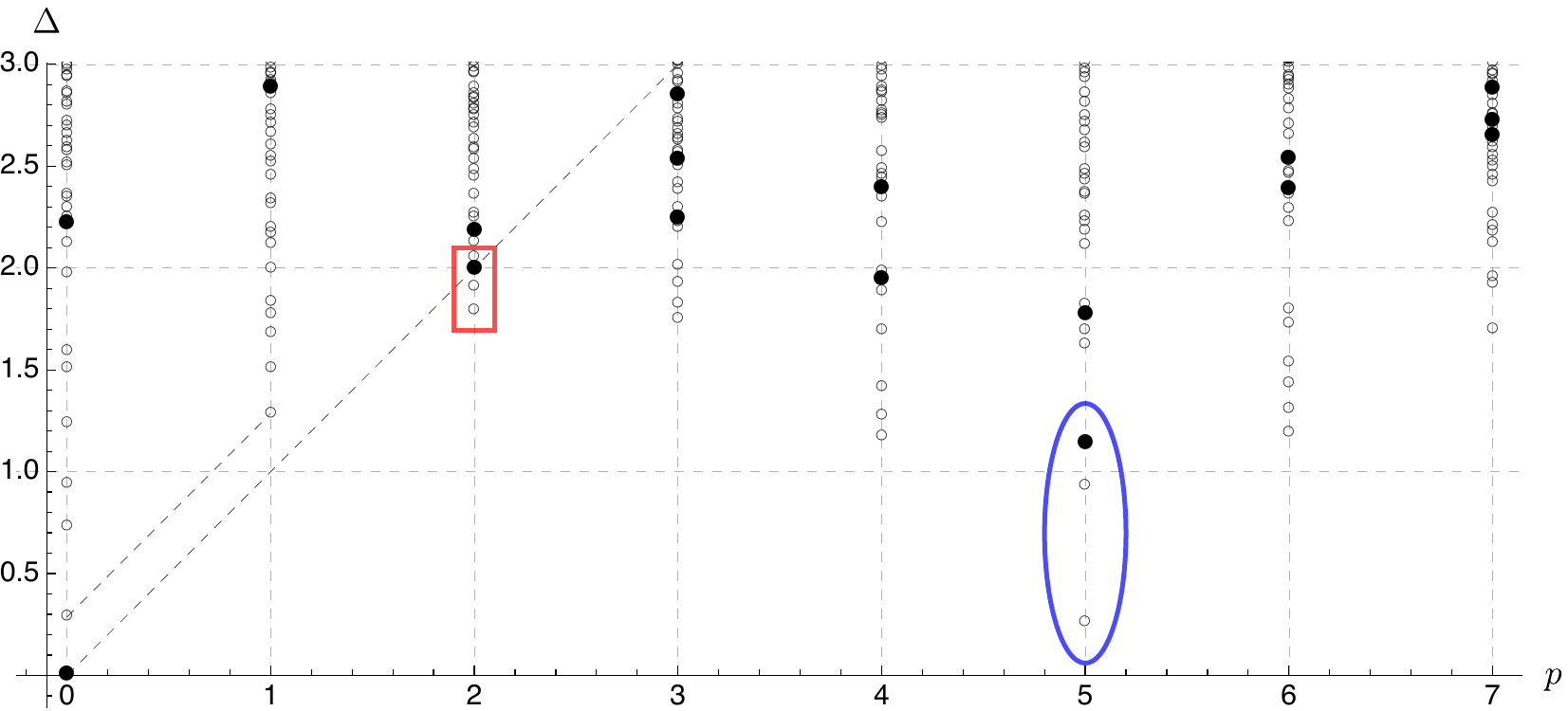}
    \\~
    \\
    \includegraphics[height=0.175\textheight]{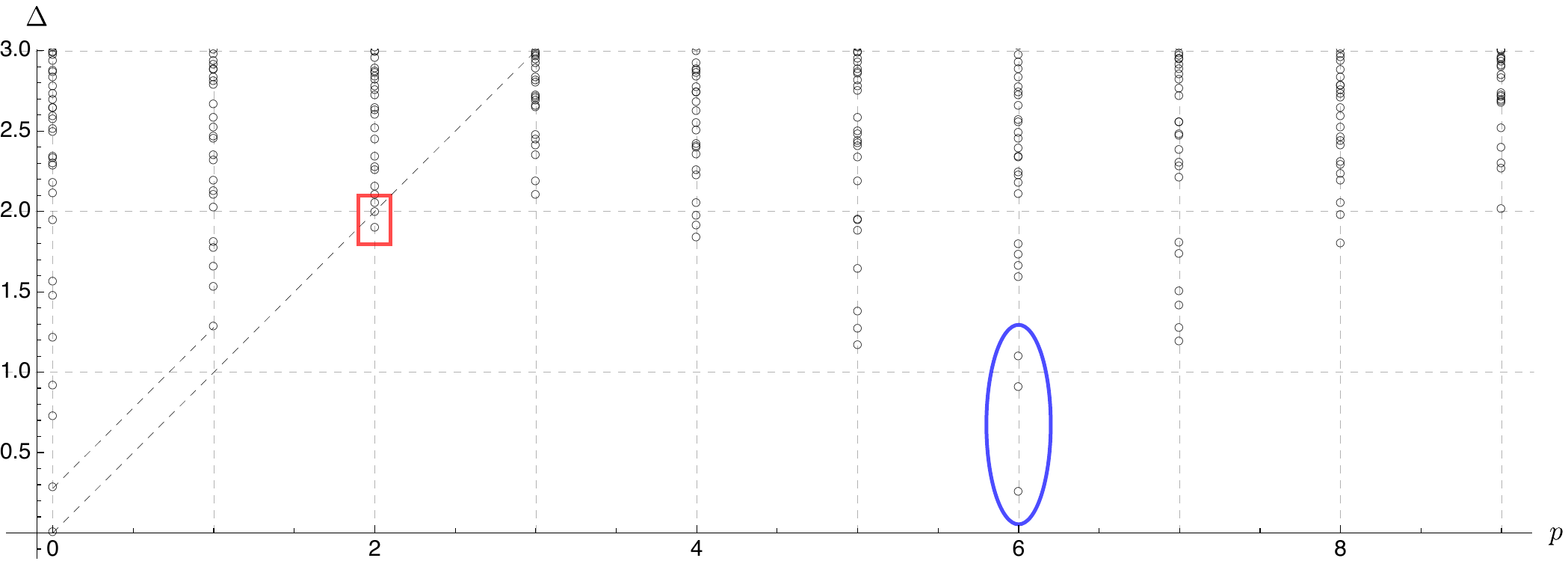}
    \caption{Periodic chain spectra for $L = 12, 15, 18$ obtained by exact diagonalization, 
    assuming that the lowest state with $p=1$ is the descendant of the first excited state with $p=0$.
    The filled dots and the hollow dots are for states with $\rho_+=\frac{3+\sqrt{13}}2$ and $\rho_-=\frac{3-\sqrt{13}}2$, respectively.  For $L = 18$ we do not yet have the $\rho$ measurement.
}
    \label{Fig:L0mod3}
\end{figure*}

Next, we present in Fig.~\ref{Fig:L0mod3} the low-lying spectra of the periodic chain obtained by exact diagonalization, where we only display states neutral under the intrinsic $\bZ_3$.
In this situation, the fusion rule for $\rho$ effectively becomes 
$\rho^2 = 1+(1+a+a^2)\rho= 1+3\rho$,
meaning that the eigenvalues of $\rho$ of states are given by $\rho_\pm=\frac{3\pm\sqrt{13}}2$.
States are represented by filled dots for $\rho_+$
and hollow dots for $\rho_-$,
which we measured according to the algorithm of \cite{Buican:2017rxc,Aasen:2020jwb}.
The horizontal axis is the momentum $p$, so that the eigenvalue under the shift of one lattice site is $e^{2\pi i p/L}$.
In Fig.~\ref{Fig:L0mod3}, we obtained $\Delta_i$ from $E_i-E_0$ 
by demanding that they satisfy $E_i-E_0 = \vvv \Delta_i/L$, where $\vvv$ was fixed by assuming that the lowest state with $p=1$ is the conformal descendant of the first excited state with $p=0$.
This normalization puts the lowest $\rho_+$ state with $p=2$ at $\Delta\sim 2$ to about 0.2\% accuracy, which strongly indicates that this state is the stress-energy tensor, and provides a consistency check of our identifications.
Assuming $c=2$, our determination of $\vvv$ translates to $\gamma=-\vvv c/12 \sim -1.0$ for the ground state of the periodic chain at $L=18$, 
which is not too distant from the value shown in Fig.~\ref{fig:energy}.

The dashed diagonal line in Fig.~\ref{Fig:L0mod3} indicates the CFT unitarity bound $\Delta \ge |p|$  in the shift-$\bZ_3$-neutral sector, so states far outside the cone should be charged under the shift-$\bZ_3$.
For example, the states within the blue ovals in Fig.~\ref{Fig:L0mod3} clearly have momentum $p=L/3$, i.e.~has the phase $e^{2\pi i/3}$ under the one-site shift.
This is reinterpreted as a shift-$\bZ_3$ charge in the continuum limit.
By contrast, the states with $p=2$ within the red rectangles appear to violate the unitarity bound, but their $\Delta$ increased as we increased $L$ from 12, 15 and then to 18.
We expect that this tendency continues as we further increase $L$, but this clearly needs further investigation.

With this understanding, the only symmetry-preserving relevant operators seen in the spectra are charged under the shift-$\bZ_3$ symmetry. 
This explains our finding that the chain is critical without any tuning.

\begin{figure*}[!ht]
    \centering
    \includegraphics[height=0.175\textheight]{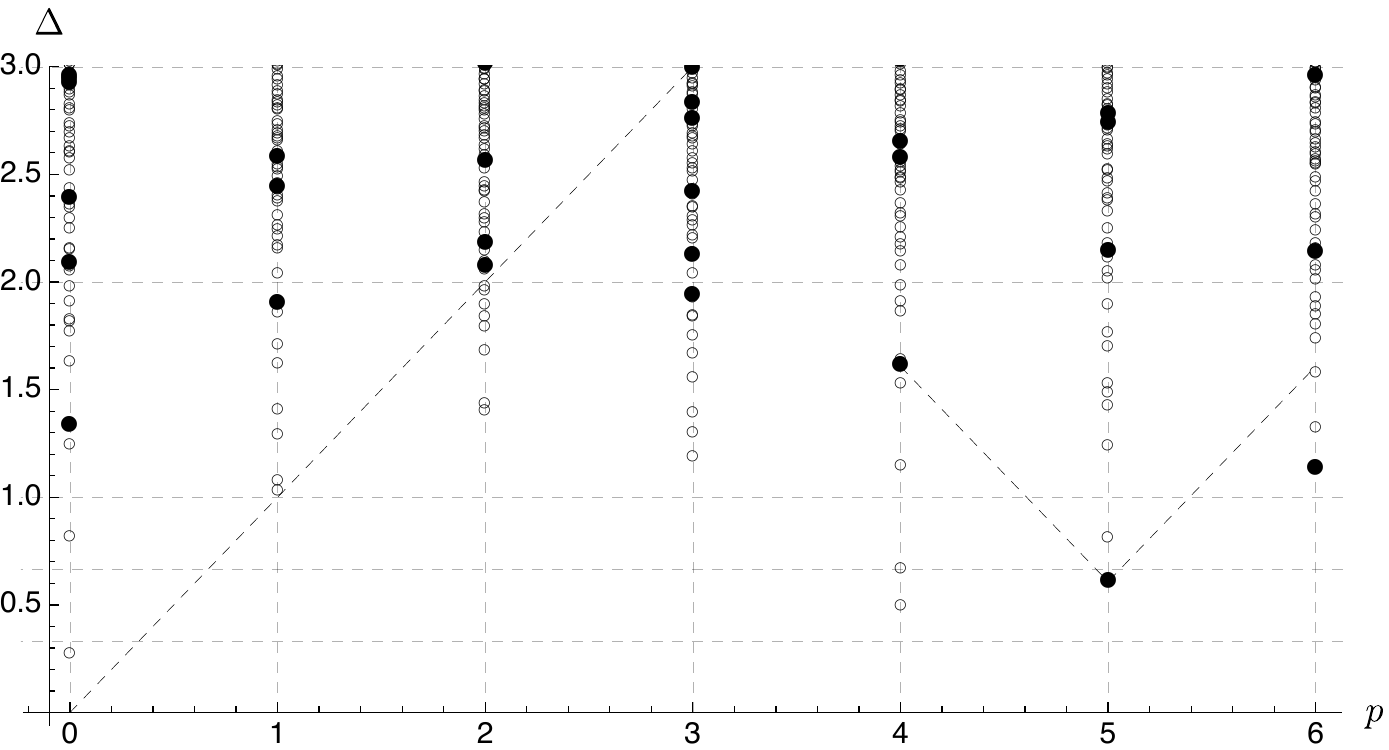}
    \includegraphics[height=0.175\textheight]{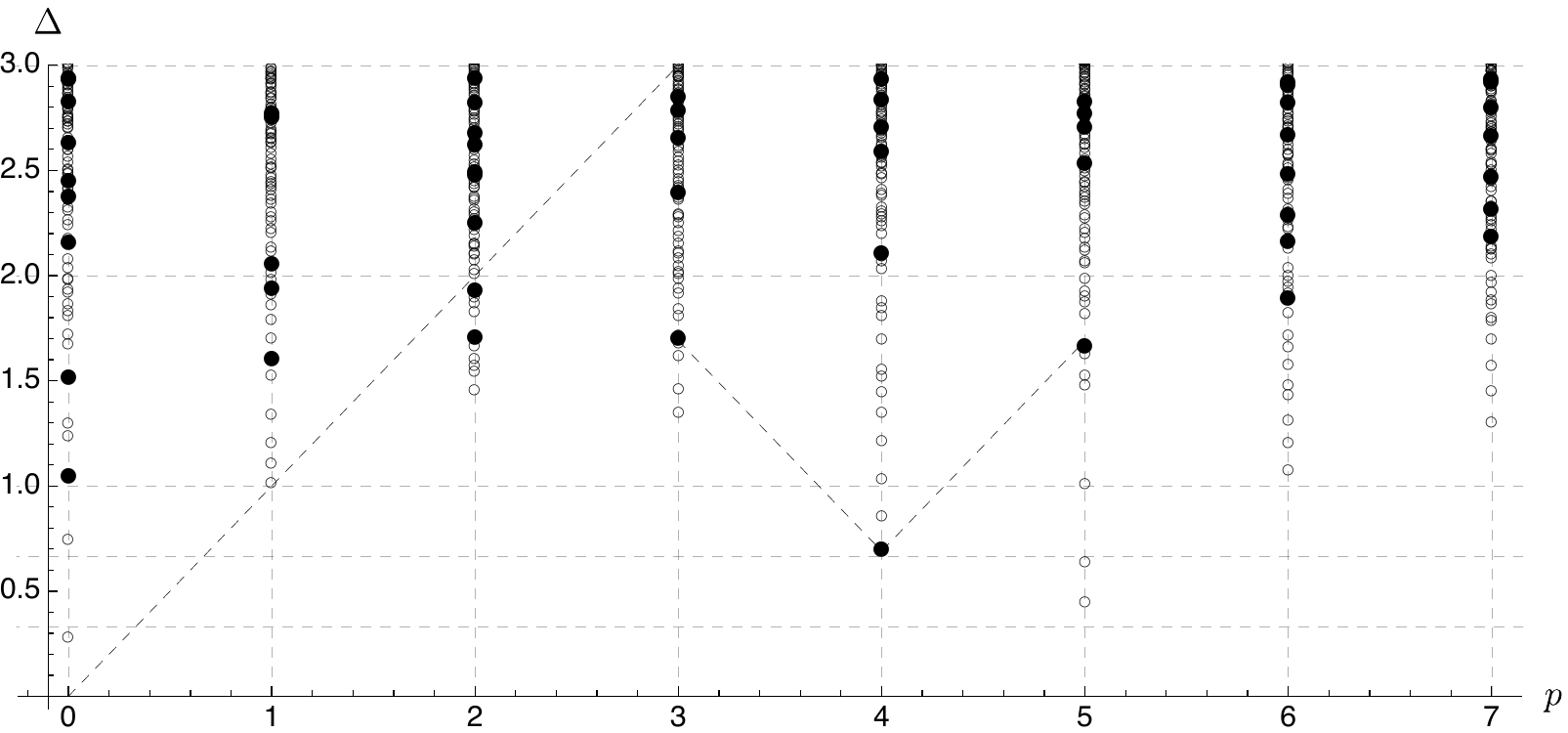}
    \caption{Periodic chain spectra for $L = 13, 14$ obtained by exact diagonalization, assuming that the lowest $\rho_+$ state with $p=(L\mp 1)/3$ is the descendant of the lowest $\rho_+$ state with $p=(L\pm2)/3$.
}
    \label{Fig:L1mod3}
\end{figure*}

Finally, we present in Fig.~\ref{Fig:L1mod3} the spectra obtained by exact diagonalization for $L=13$ and $14$, which should correspond to the sectors twisted by the shift-$\bZ_3$ symmetry.
We fixed $v$ by demanding that the lowest $\rho_+$ state at $p=(L\mp1)/3$ is the conformal descendant of the lowest $\rho_+$ state at $p=(L\pm2)/3$.
This determines $\Delta_i$ using \eqref{generalFit} with the known value of $\alpha$ and $c=2$.
We see a $\rho_-$ operator with $p=1$ and $\Delta\sim 1$.
If it truly saturates the unitarity bound $\Delta\ge |p|$, then it would mean that there is a spin-1 conserved current in the shift-$\bZ_3$ twisted sector, which would in turn imply that our model is a $\bZ_3$-orbifold of a sigma model on $T^2$.

\newpage

\begin{acknowledgments}
\textit{Acknowledgements.---}The authors thank
Minjae Cho, Chang-Tse Hsieh, Yifan Wang, and Xi Yin for helpful discussions;
Chang-Tse Hsieh, Justin Kaidi, Hosho Katsura, Yunqin Zheng for comments on the draft;
Masahiko G. Yamada for  discussions on numerical algorithms; Matthew Fishman and Jie Ren for consultation through the iTensor Support Q\&A;
Gen Kuroki, Hosho Katsura, and Kensuke Tamura for help on setting up the Julia code;
and Paul Fendley, Alexander Hahn, Laurens Lootens, Tobias J. Osborne, Robijn Vanhove, Frank Verstraete, and Ramona Wolf for correspondences.
Finally, YT thanks Yasuyuki Kawahigashi for first introducing him to the Haagerup subfactor a long time ago, which was the ultimate source of this project.

The numerical computations were performed on the Harvard FAS Research Computing cluster and the IPMU iDark cluster.

T.H.~is supported by the U.S. Department of Energy, Office of Science, Office of High Energy
Physics, under Award Number DE-SC0011632.
Y.L.~is supported by the Simons Collaboration Grant on the Non-Perturbative Bootstrap.
Y.T.~is partially supported  by JSPS KAKENHI Grant-in-Aid (Wakate-A), No.17H04837
and also by WPI Initiative, MEXT, Japan at IPMU, the University of Tokyo.
M.T.~is partially supported by JSPS KAKENHI No. JP17K17822, JP20K03787, JP20H05270, and JP21H05185. 

The authors of this paper were ordered alphabetically.
\end{acknowledgments}


\bibliographystyle{ytphys}
\let\raggedright\relax
\bibliography{ref}

\end{document}